\documentclass[12pt]{iopart}
\usepackage{epsf,epsfig,latexsym}
\usepackage{bm}
\usepackage{graphicx}
\usepackage{multirow}
\usepackage[rflt]{floatflt}
\usepackage{units}
\DeclareRobustCommand{\dgr}[1][]{
  \unit[#1]{\ifmmode{}^\circ\else${}^\circ$\fi}}

\def\ssNN#1{\sqrt{s_{NN}} \ifx|#1|\else=\unit[#1]{GeV}\fi}
\DeclareRobustCommand{\snn}[1]{\ifmmode\ssNN{#1}\else$\ssNN{#1}$\fi}
\DeclareRobustCommand{\valn}[3]{\ifmmode #1\,\pm\,_{#2\text{ (stat)}}^{#3\text{ (syst)}}\else$#1\,\pm\,_{#2\text{ (stat)}}^{#3\text{ (syst)}}$\fi}
\DeclareRobustCommand{\val}[3]{\ifmmode #1\,\pm\,_{#2}^{#3}\else$#1\,\pm\,_{#2}^{#3}$\fi}

\DeclareRobustCommand{\mpt}
{\ifmmode\left<p_T\right>\else$\left<p_T\right>$ \fi}
\DeclareRobustCommand{\pT}{\ifmmode p_T\else$p_T$\fi}
\DeclareRobustCommand{\mTm}{\ifmmode m_T-m\else$m_T-m$\fi}

\begin{document}

\title[Do thermal parameters depend upon rapidity?]{Is there more than one thermal source?}

\author{Michael Murray for the BRAHMS Collaboration}
\address{University of Kansas, mjmurray@ku.edu, 785 864 3949}

\begin{abstract}
BRAHMS has the ability to study relativistic heavy ion collisions over a wide range of $p_T$ and rapidity \cite{BrNim}. 
This allows us to test whether thermal 
models can be generalized to describe the rapidity dependence of particle ratios. This appears to work with the baryo-chemical potential changing more rapidly than the temperature. Using fits to BRAHMS data for the 5\% most
central Au+Au collisions we are able to 
describe $\Xi$ and $\Omega$ ratios from other experiments. This paper is dedicated to Julia Thompson who worked to bring South African teachers  into physics.

\end{abstract}



\section{Introduction}
The purpose of RHIC is to map the phase structure of QCD. 
So far the community has concentrated on $AuAu$, $dAu$ and $pp$ collisions at 
$\sqrt{s_{NN}}=200$ GeV 
in the hope of finding 
the quark gluon plasma. 
BRAHMS' special contribution 
has been 
to study the hadrons produced in these collisions over a broad range of 
$p_T$ and rapidity.  
The distribution of particles in rapidity and $p_T$ may give information on the transverse and longitudinal flow while the mix of different kinds of particles may tell us about the  ``quark chemistry" of the system. It is on this aspect of the bulk behavior that we shall concentrate in this paper. 

\section{Particle Yields and Transverse Momenta}

\begin{figure}[hbt]
 \begin{minipage}[b]{.45\textwidth}
  \includegraphics[width=\columnwidth]{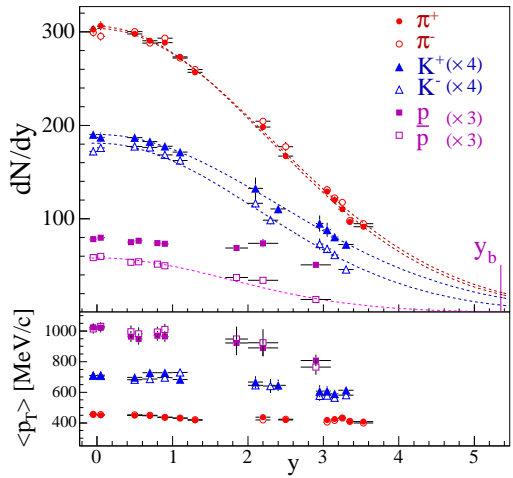}
  \caption{Rapidity densities (top) and  mean transverse momentum (bottom)
as a function of rapidity. The lines show Gaussian fits.}
  \label{dndy}
  \end{minipage}
\end{figure}

BRAHMS has  measured particle spectra over a very wide range of rapidity and  $p_T$.
Our $AuAu$ spectra are summarized in Fig.~\ref{dndy}, which shows the 
rapidity densities, dN/dy, and 
the mean transverse momenta, \mpt{}, for $\pi, K$, p and $\bar p$
 as a function of rapidity \cite{BrMeson}.  
For $\pi, k$ and $\bar p$ 
the yields peak at y=0 and drop significantly at higher
rapidities.  The $\pi^+$ and $\pi^-$ yields are nearly equal
while an excess of $K^+$ over $K^-$ is observed that increases with rapidity.

The mean $p_T$s, \mpt{}, for all particles decline slowly with rapidity. We have fitted the spectra themselves to the ``blast wave" model in order to estimate the transverse flow and kinetic freeze-out temperature \cite{blast}. We find that the flow decreases with rapidity while the kinetic temperature tends to increase \cite{MurrayQM04}.

\section{Rapidity Dependence of  Chemical Freeze-out}

Traditionally thermal analyses have assumed that there is only one source in heavy ion collisions and have used particle yields integrated over all phase space as input. 
At RHIC several groups have shown that such models can give an excellent description of a large number of particle ratios at {\it mid-rapidity} 
The use of mid-rapidity ratios has been justified on the grounds that there is a boost invariant region near y=0. Such a hypothesis implies that there is a different source away from mid-rapidity. 
We have investigated this idea by fitting our  particle yields 
at several different rapidities to a chemical model 
\cite{Becattini}. 

\begin{figure}[ht]
 \begin{minipage}[b]{3.0in}
  \includegraphics[width=3.0in]{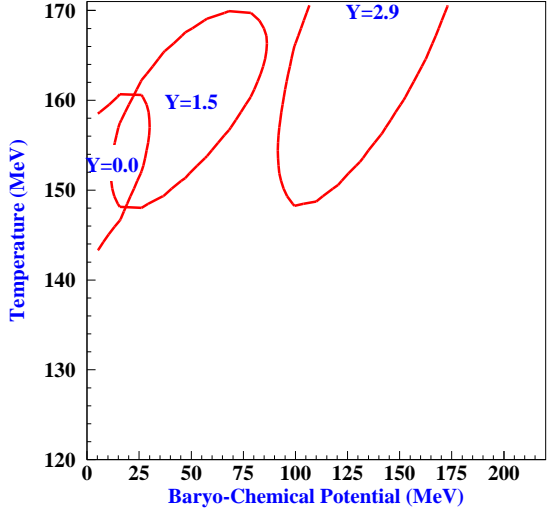}
  \caption{Preliminary thermal (right) fits to BRAHMS $AuAu$ data at various rapidities. The curves are the one sigma contours.}
  \label{TvMuB}
   \end{minipage}
\end{figure}

\begin{figure}[ht]
 \begin{minipage}[b]{3.0in}
  \includegraphics[width=3.0in]{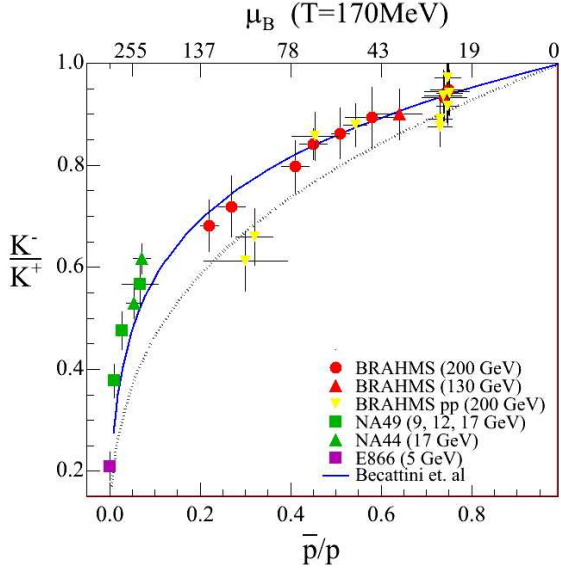}
  \caption{Correlation between $K^-/K^+$ and  $\bar p/p$ ratios for central $AuAu$ and $pp$ collisions. The NA44 and NA49 the results are for central PbPb collisions.}
  \label{KmpVpmp}
   \end{minipage}
\end{figure}

Figure~\ref{TvMuB} shows  
that as y increases both the baryo-chemical potential and (to a lesser extent) the chemical freeze-out temperature increase. This may suggest that the system has fewer degrees of freedom at higher rapidities.  
Figure~\ref{TvMuB} suggests that chemical freeze-out temperature changes more slowly with rapidity than the baryo-chemical potential. It is possible that this is also true as we change the collision energy? 
Figure~\ref{KmpVpmp} shows the correlation between 
 $K^-/K^+$ and  $\bar p/p$ ratios measured by BRAHMS for $AuAu$ and $pp$ collisions at several rapidities. Lower energy results are also shown. Both  the  $AuAu$ and $pp$ results can be described by a power law,
 $K^-/K^+ = ({\bar p}/p)^\alpha$ with $\alpha = 0.32 \pm 0.4$ for $pp$ and 
$0.28 \pm 0.06$ for $AuAu$ \cite{BrMeson,BrPpRatios}. Preliminary $AuAu$ data from $\sqrt{s_{NN}} = 63$ GeV are also consistent with this fit \cite{DjamelSqm04}. 

\begin{figure}
\begin{minipage}[b]{3.5in}
  \includegraphics[width=3.5in]{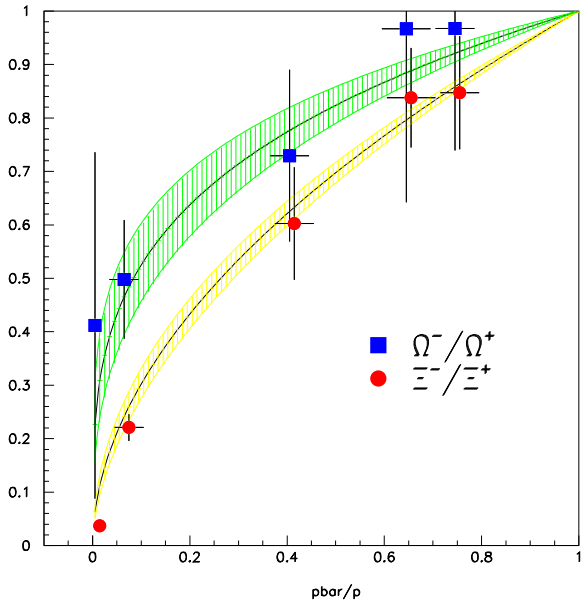}
  \caption{${\bar \Omega}/\Omega$ and 
${\bar \Xi}/\Xi$ ratios from STAR and NA57 versus predictions Eqns 
[1] and [2] \cite{NA57Hyperons,StarHyperons}. The 63 and 200 GeV data are still preliminary. The ${\bar p}/p$ ratios have been displaced slightly for clarity. The bands represent the error on the determination of $\mu_s$ from the BRAHMS data.}
  \label{XiOmega}
  \end{minipage}
\end{figure}

Recasting the power law fit in terms of chemical potentials gives $\mu_s = (0.28 \pm 0.6) \cdot \mu_q$ independent of temperature. If we take this seriously then we should be able to predict the correlation between other strange particle ratios and $\bar p/p=e^{-6\mu_q/T}$ just by counting the number of strange and non-strange quarks. For example
\begin{equation}
   \frac{\bar \Omega}{\Omega} = e^{-6\mu_s/T}=
e^{-0.28\cdot 6\mu_q/T}=\left(\frac{\bar p}{p} \right)^{0.28}
\label{eq:Omega}
\end{equation}
and 
\begin{equation}
   \frac{\bar \Xi}{\Xi} = e^{-(4\mu_s+2\mu_q)/T}=
e^{-(1.12 +2)\cdot \mu_q/T}=\left(\frac{\bar p}{p} \right)^{0.52}.
\label{eq:Xi}
\end{equation}
Figure~\ref{XiOmega} shows the predictions of Equations (1) and (2) as well 
as preliminary $\Xi$ and $\Omega$ ratios from NA57 ($\sqrt{S_{NN}}=$9 and 17GeV) and STAR ($\sqrt{S_{NN}}=$63, 130 and 200GeV) \cite{NA57Hyperons,StarHyperons}. The agreement between the prediction and data is  very good.  Note that our ignorance of the temperature T does not affect the shape of these curves.

\section{Summary and Conclusions} 

Thermal descriptions are a powerful way to describe $AuAu$ collisions.
Chemical analysis of our particle yields hint that both the baryo-chemical potential and the chemical freeze-out temperature increase with rapidity. 
One could interpret this in terms of the system becoming less partonic (with consequently fewer degrees of freedom) at higher rapidities. 
At this stage however the data are also consistent with a constant temperature at different rapidities. The change of baryo-chemical potential seems to dominate the rapidity (and energy) dependence of the particle ratios. However the correlation between $K^-/K^+$ and  $\bar p/p$ is different for $pp$ and heavy ion collisions. This may be an effect of the small size of the $pp$ system. For central $AuAu$ ($PbPb$) collisions the relationship 
$\mu_s = (0.28 \pm 0.6) \cdot \mu_q$ derived from BRAHMS data gives a good description of hyperon ratios.

\section*{References}
\bibliographystyle{unsrt}

\end{document}